\documentclass{mn2e}

\usepackage[dvips]{graphicx}
\usepackage{amssymb}
\usepackage{txfonts}

\newcommand{\sax}{{\it Beppo\-SAX}}
\newcommand{\integral}{{\it INTEGRAL}}
\newcommand{\msun}{{\rm M}_{\sun}}
\newcommand{\xte}{{\textit RXTE}}

\title[X-ray and radio states of Cyg X-3]{A classification of the X-ray and radio states of Cyg X-3 and their long-term correlations}

\author[A. Szostek, A. A. Zdziarski and M. L. McCollough]
{
Anna Szostek,$^{1,2}$\thanks{E-mail: asz@camk.edu.pl, aaz@camk.edu.pl}
Andrzej A.~Zdziarski$^{2}$\footnotemark[1] and 
Michael L.~McCollough$^{3}$\\
$^{1}$ Astronomical Observatory, Jagiellonian University, Orla 171, 30-244
Krak\'ow, Poland\\
$^{2}$ Centrum Astronomiczne im.\ M. Kopernika, Bartycka 18, 00-716 Warszawa, Poland\\
$^{3}$ Smithsonian Astrophysical Observatory, 60 Garden Street, MS 67, Cambridge, MA 02138-1516, USA\\
}

\date{Accepted 2008 May 16. Received 2008 April 28; in original form 2008 March 14}

\pagerange{\pageref{firstpage}--\pageref{lastpage}}
\pubyear{2008}

\begin{document}

\maketitle

\label{firstpage}

\begin{abstract}
We present a detailed classification of the X-ray states of Cyg X-3 based on the spectral shape and a new classification of the radio states based on the long-term correlated behaviour of the radio and soft X-ray light curves. We find a sequence of correlations, starting with a positive correlation between the radio and soft X-ray fluxes in the hard spectral state, changing to a negative one at the transition to soft spectral states. The temporal evolution can be in either direction on that sequence, unless the source goes into a very weak radio state, from which it can return only following a major radio flare. The flare decline is via relatively bright radio states, which results in a hysteresis loop on the flux-flux diagram. We also study the hard X-ray light curve, and find its overall anticorrelation with the soft X-rays. During major radio flares, the radio flux responds exponentially to the level of a hard X-ray high-energy tail. We also specify the detailed correspondence between the radio states and the X-ray spectral states. We compare our results to those of black-hole and neutron-star binaries. Except for the effect of strong absorption and the energy of the high-energy break in the hard state, the X-ray spectral states of Cyg X-3 closely correspond to the canonical X-ray states of black-hole binaries. Also, the radio/X-ray correlation closely corresponds to that found in black-hole binaries, but it significantly differs from that in neutron-star binaries. Overall, our results strongly support the presence of a black hole in Cyg X-3. 
\end{abstract}

\section{Introduction}
\label{intro}

Cyg X-3 is the brightest radio source among X-ray binaries (McCollough et al.\
1999). It is a high-mass system with a Wolf-Rayet companion (van Kerkwijk et al.\ 1996), but with an unusually short orbital period, $P=4.8$ h. It is located at a distance of $d\simeq 9$ kpc in the Galactic plane (Dickey 1983; Predehl et al.\ 2000). In spite of its discovery in 1966 (Giacconi et al.\ 1967), the system remains poorly understood. In particular, due to the lack of a reliable mass function, it remains uncertain whether its compact object is a black hole or a neutron star. Currently, there are two other known X-ray binaries containing Wolf-Rayet stars, IC 10 X-1 (Prestwich et al.\ 2007) and NGC 300 X-1 (Carpano et al.\ 2007a, b). In the former case, there is a very strong dynamical evidence that the compact object is a black hole of a high mass, $\ga 30\msun$ if the companion has the mass indicated by its spectrum, and $\ga 20\msun$ for the minimum possible companion mass (Silverman \& Filippenko 2008). Based on the luminosities observed from Cyg X-3 in its different spectral states, Hjalmarsdotter et al.\ (2008a, hereafter Hj08a) and Szostek \& Zdziarski (2008, hereafter SZ08) similarly found the most likely mass of the compact object to be $\ga 20\msun$. In the case of NGC 300 X-1, Carpano et al.\ (2007a) argued for the presence of a black hole based on its very high luminosity and the similarity of the system to IC 10 X-1. Thus, it is very interesting that the three currently known Wolf-Rayet X-ray binaries may all contain black holes. 

As an X-ray source, Cyg X-3 is persistent and observed in two main spectral states, hard and soft. Recently, the hard state has been studied in detail by Hj08a, and both the hard and soft states, by SZ08. However, there have yet been no comprehensive studies of the range of the states including the transitionary ones. Interestingly, the spectra of Cyg X-3 in spite of being generally similar to the canonical states of black hole binaries show some significant differences (Zdziarski \& Gierli\'nski 2004, hereafter ZG04). Their physical interpretation also remains rather uncertain (Hj08a; SZ08). A major issue is the strong and complex intrinsic absorption, most likely caused by the wind from the companion star, which greatly complicates the determination of the intrinsic spectral shapes and luminosities. SZ08 have studied in detail the absorption in the Wolf-Rayet stellar wind, including effects of its velocity profile, the elemental composition, irradiation by the X-ray source, and inhomogeneities. They were able to determine the wind mass-loss rate as $\dot M\sim 10^{-5}\msun\, {\rm yr}^{-1}$. However, even that detailed treatment of the wind absorption did not allow for an unambiguous determination of details of the underlying radiative processes in the two main spectral states. Thus, it is highly desirable to further study the X-ray states of this system.

In the radio, Cyg X-3 is a persistent source in the sense that is always detected, though its flux varies by four orders of magnitude, as well as the relative contributions of the core and jets vary strongly (e.g., Tudose et al.\ 2007). On the basis of 1988--1992 radio observations by the Green Bank Interferometer (GBI), Waltman et al.\ (1994, 1995, 1996) identified four of its radio states. They include the so-called quiescent radio emission, with a $\sim$50--200 mJy flux variable on a timescale of months, episodes of frequent minor-flaring, $\la 0.3$ Jy, very weak, or quenched, emission, $\la 30$ mJy, and major flares, $\sim$1--20 Jy, which follow the quenched state. (Note that the term 'quiescent' used here is not related to the so-called quiescent state of low-mass X-ray binaries, which corresponds to much weaker states than the above one of Cyg X-3.) Interestingly, all major radio flares were preceded by very low radio flux levels. However, apart from that sequence, those authors did not study either the temporal patterns of the radio states or their connection to the X-ray states.

The existence of a relationship between the two energy bands was first pointed out by Watanabe et al.\ (1994) based on simultaneous monitoring by the GBI and the All Sky Monitor (ASM) on board of {\it Ginga}. They showed that strong radio flares occur only when the source was in its soft state. Then, McCollough et al.\ (1999) studied the relationship between the radio emission measured by the GBI and the hard X-ray, 20--100 keV, emission measured by the Burst and Transient Source Experiment (BATSE) on board of {\it Compton Gamma Ray Observatory}. They found an anticorrelation in the quiescent state (corresponding to the hard state in X-rays), a positive correlation during major radio flares and in the quenched state, and no correlation during minor flares. Gallo, Fender \& Pooley (2003, hereafter GFP03) and Hj08a have presented the correlation between the 1.5--12 keV X-ray flux measured by the ASM on board of {\it Rossi X-ray Timing Explorer\/} (\xte) and the 15 GHz measurements by the Ryle telescope. They found a clear positive correlation in the hard X-ray state, and its break down above certain X-ray flux. The sign of the hard-state correlation of the 1.5--12 keV flux is opposite to that of the 20--100 keV flux, which is consistent with the anticorrelation between those two X-ray bands (e.g., Choudhury et al.\ 2002; Hj08a). However, details of the radio/X-ray behaviour in the soft state were not clear from these works. 

In this paper, we present our new classifications of the X-ray and radio states of Cyg X-3, and study the long-term radio/X-ray correlation including the transitionary and soft states. In Section \ref{xray}, we present a detailed classification of the X-ray spectral states based on the X-ray data from the \xte\/ Proportional Counter Array (PCA) and the High Energy X-Ray Transient Experiment (HEXTE) (a preliminary account of some of these results was given in Szostek \& Zdziarski 2004). In Section \ref{radio}, we re-classify the radio states based the long-term temporal behaviour of the GBI (radio) and \xte\/ ASM (soft X-ray) light curves, and relate these states to the X-ray spectral states. Also, we analyze correlations of hard X-rays with the other bands using data from the BATSE. In Section \ref{discussion}, we summarize and discuss our results, as well as compare them to the corresponding results for black-hole and neutron-star X-ray binaries. 

\section{The data set}
\label{data}

In our analysis, we use the 3--5 keV soft X-ray band of the \xte\/ ASM dwell data. As we have found out, the use of this band yields the cleanest radio/X-ray correlation. On the other hand, the ASM 5--12 keV band overlaps with the region where the Cyg X-3 spectra pivot, which results in the count rate of spectra with different shape being similar, which then smears the correlation. Then, the 1.5--3 keV band is very strongly absorbed with the absorbing column being variable, which also results in a smearing of the correlation. Also, the 3--5 keV band has the narrowest relative width among the three, and thus its count rate is closest to a monochromatic flux. 

We then use the 20--100 keV hard X-ray data from the BATSE. The data have been obtained using an Earth occultation analysis technique (Harmon et al.\ 2002); see McCollough et al.\ (1999) for analysis of an earlier part of the BATSE data. Each point represents a 3-day average of the energy flux.

The radio data have been collected during the monitoring program of the GBI (see Waltman et al.\ 1994, 1995, 1996). Here, we use the data for MJD 50409--51823, during which period observations of Cyg X-3 were performed several times per day on most days. For our analysis we use the 8.3 GHz data. This choice is arbitrary, the use of either the GBI 2.25 GHz data or the 15 GHz Ryle Telescope data yields relatively similar results. 

We then rebin the ASM and GBI light curves in such way that the significance of a single bin is $\geq 3 \sigma$. The rebinned data are also plotted in the radio/soft X-ray correlation diagrams shown in Section \ref{radio} below requiring each pair of points to be taken on the same day. Given the relatively low statistical quality of the BATSE data, we rebin its light curve to $\geq 8\sigma$ significance (but requiring the distance between the midpoints of the data in a given bin to be $\leq 10$ days). In showing correlations of the hard X-ray data with the radio and soft X-rays, we slightly modify this criterion in order to achieve the best clarity of the graphical representations. Also, since the measured fluxes from the BATSE are often null, we use a linear scale for it (while we use logarithmic scales for all other data).

We also use selected \xte\/ PCA and HEXTE pointed observations. We have selected 42 pointed observations from 1996--2000 for which the PCA data from the Proportional Counter Units 0--2 were available. We have extracted the PCA spectra in the 3--25 keV range using the top layer of the detectors. We then obtained the corresponding HEXTE spectra, using both HEXTE clusters, in the 15--110 keV energy band. In our spectral fits, we allow a free relative normalization of the HEXTE spectra with respect to that from the PCA.

\section{The X-ray spectral states}
\label{xray}

\begin{figure}
\centerline{\includegraphics[width=\columnwidth]{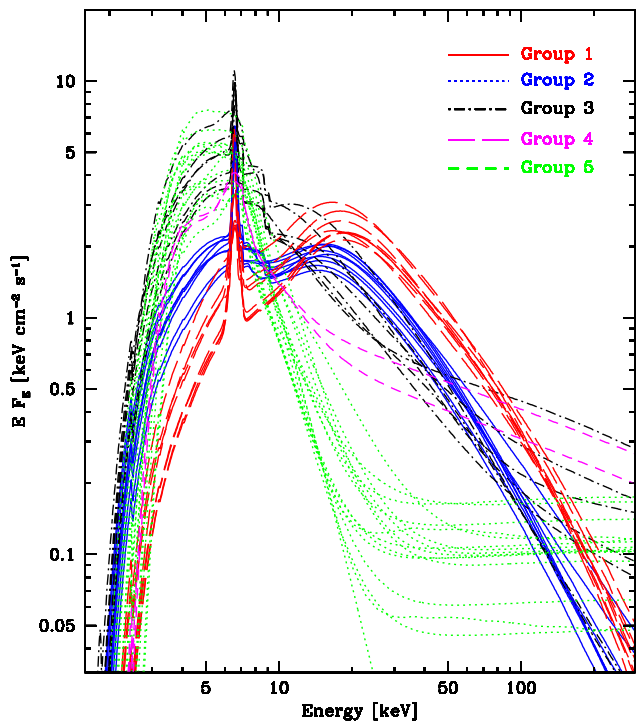}}
\caption{Comptonization model spectra of Cyg X-3 from the 42 pointed \xte\/
observations. Different line styles and colours correspond to our classification of the spectra (the same colours are also used in Figs.\ \ref{groups} and \ref{asmradio_pointed} below). The extreme states contain spectra with a strong soft X-ray emission followed by a weak hard X-ray tail (group 5), and those with a weak soft X-ray emission and hard X-rays peaking around $\sim$20 keV (group 1).} 
\label{individual}
\end{figure}

\begin{figure}
\centerline{\includegraphics[width=\columnwidth]{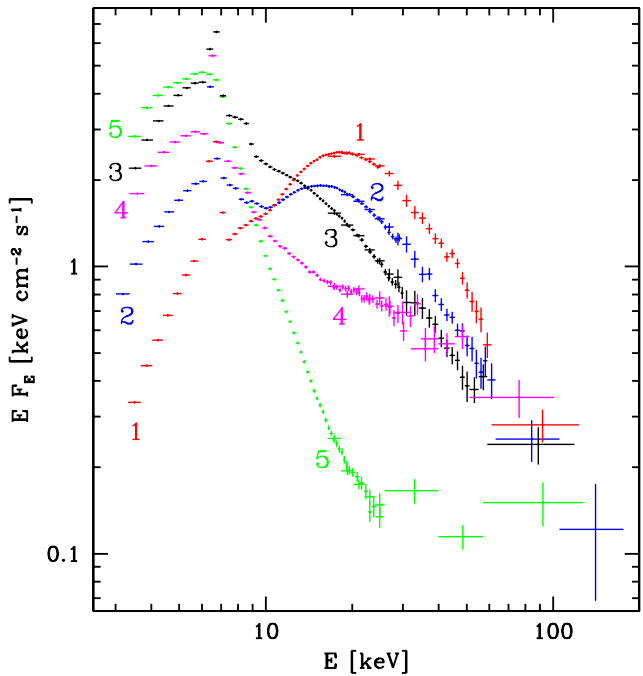}}
\caption{Deconvolved spectra for the average of each of the 5 groups shown on Fig.\ \ref{individual}, fitted with the same model as the individual spectra. The HEXTE data are renormalized to the level of the PCA.}
\label{groups}
\end{figure}

We fit each of the 42 PCA/HEXTE spectra with the same model as applied to \integral/\xte\/ spectra from Cyg X-3 by Vilhu et al.\ (2003). Similar models were also used by Hj08a and SZ08. The model includes Comptonization by hybrid (i.e., both thermal and nonthermal) electrons (Coppi 1999, Gierli\'nski et al.\ 1999), Compton reflection from an ionized medium (Magdziarz \& Zdziarski 1995), absorption by fully and partially covering neutral media, and a Gaussian Fe K fluorescent line. As discussed in Vilhu et al.\ (2003), this model treats the low-energy part of the spectrum only phenomenologically. Still, it provides good fits to the PCA/HEXTE spectra and a physical description of the hard X-rays. 

Furthermore, a sophisticated stellar wind absorption model was used by SZ08 in fits to X-ray spectra from two \sax\/ observations, in the hard and soft state. Surprisingly, it was found that the model yields overall intrinsic spectral continua roughly similar to those obtained when the same data were fitted by the above phenomenological model, compare figs.\ 6 and 11 in SZ08. Therefore, though the fitted absorber parameters of the phenomenological model cannot be considered as a realistic description of the absorber, the resulting spectral shapes may, in principle, yield a fair approximation to the actual intrinsic X-ray spectra. However, details of those fits are beyond the scope of this work and are presented elsewhere (Hjalmarsdotter et al.\ 2008b). 

The obtained (absorbed) model spectra are shown in Fig.\ \ref{individual}. We have divided them into five groups based on their spectral shape and ordered by the decreasing flux at 20 keV. The increasing group number also roughly corresponds to the decreasing spectral hardness in the 10--20 keV range. The groups 1--2, 3--4, and 5 may be classified as belonging to the hard, intermediate/soft and ultrasoft state, respectively. Some of the flux variability within each group is caused by the orbital modulation. 

\begin{table}
\caption{The average (corresponding to the spectra in Fig.\ \ref{groups}) {\it absorbed\/} bolometric X-ray fluxes, $F_{\rm X}$, and isotropic luminosities, $L_{\rm X}$ (for $d=9$ kpc), for the 5 spectral types. }
\begin{center}
\begin{tabular}{ccc}
\hline
Type & $F_{\rm X}$                    & $L_{\rm X}$ \\
& (erg cm$^{-2}$ s$^{-1}$) & (erg s$^{-1}$) \\
\hline
1 & $7.8   \times 10^{-9}$ & $7.5 \times 10^{37}$ \\
2 & $7.7   \times 10^{-9}$ & $7.5 \times 10^{37}$ \\
3 & $1.1   \times 10^{-8}$ & $1.1 \times 10^{38}$ \\
4 & $8.2   \times 10^{-9}$ & $7.9 \times 10^{37}$ \\
5 & $8.1   \times 10^{-9}$ & $7.9 \times 10^{37}$ \\
\hline
\end{tabular}
\end{center}
\label{fluxes}
\end{table}

\begin{figure*}
\centerline{\includegraphics[height=175mm,angle=-90]{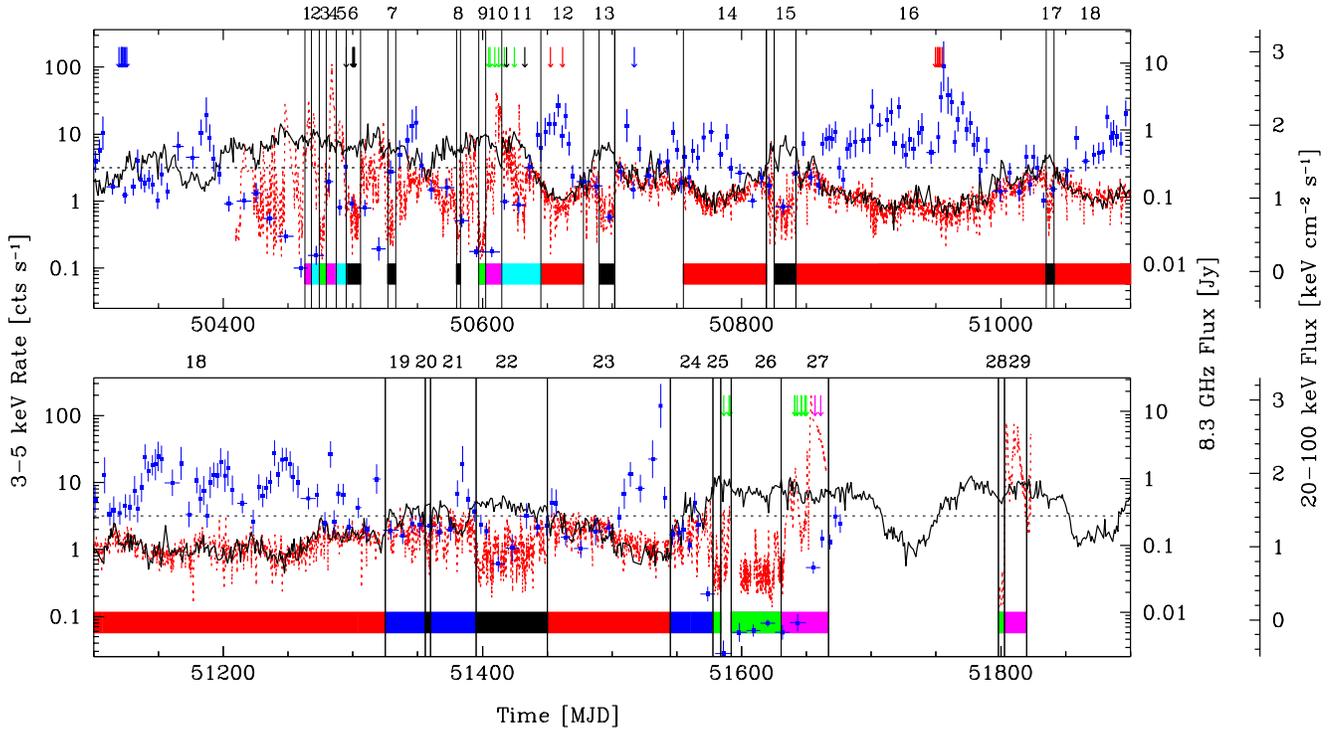}}
\caption{The \xte\/ ASM 3--5 keV, BATSE 20--100 keV, and GBI 8.3 GHz light curves, shown as the black solid curves, blue crosses, and red dotted curves, respectively. The horizontal dotted line represents the transition level at which the radio/soft X-ray correlation changes its character. The vertical lines divide the light curves into intervals of different activity types. The numbers on top of the intervals are listed in Table \ref{radio_states}, where patterns of similar behaviour are grouped together, and identified by the colours of the boxes marking the numbered intervals. Note that some intervals have remained unclassified. The arrows show the times of the pointed PCA/HEXTE observations, and their colours identify the spectral states shown in Figs.\ \ref{individual}--\ref{groups}.
}
\label{asmradio}
\end{figure*}

We have then created the average PCA/HEXTE spectra within each group. We have fitted them with the same model as above, obtaining $0.3<\chi^2_\nu< 1.9$ (with a 1 per cent systematic error added in quadrature to the statistical error). The resulting spectra are shown in Fig.\ \ref{groups}. The Cyg X-3 hard-state spectra from \xte\/ and \integral\/ of Vilhu et al.\ (2003), the \sax\/ spectra of SZ08 and the \integral\/ spectra of Hj08a are relatively close to our spectral types 1--2. The soft-state spectrum from \sax\/ of SZ08 is similar to that of our type 5. For reference, the bolometric absorbed fluxes and isotropic luminosities are given in Table \ref{fluxes}. Note that the effect of the absorption on the bolometric flux is much stronger for soft spectra (where the bolometric correction can be a factor of several or more) than for the hard ones, and thus the fluxes of the intrinsic, unabsorbed, spectra would have a significantly larger spread and be ranked differently than the absorbed fluxes given here (see Hjalmarsdotter et al., in preparation). For comparison, the absorbed and unabsorbed bolometric fluxes in the hard state measured by \sax\/ (SZ08) are $\simeq 6.5\times 10^{-9}$ erg cm$^{-2}$ s$^{-1}$ (similar to the values in Table \ref{fluxes} for the types 1--2) and $\simeq 8.5\times 10^{-9}$ erg cm$^{-2}$ s$^{-1}$, respectively, i.e., the correction for absorption is only moderate\footnote{Note that due to an unfortunate error, the fluxes given in Table 3 of SZ08 for the hard state are the absorbed ones. The unabsorbed fluxes of $F_{\rm X}$, $F_{\rm X,0}$ for the minimum and and maximum accretion rate are: 8.50, 4.28, 8.42, 4.23, respectively, in units of $10^{-9}$ erg cm$^{-2}$ s$^{-1}$.}. On the other hand, in the soft state, those fluxes are $\simeq 8.4\times 10^{-9}$ erg cm$^{-2}$ s$^{-1}$ (also similar to the value in Table \ref{fluxes} for the type 5), $\sim\! 3\times 10^{-8}$ erg cm$^{-2}$ s$^{-1}$, respectively. Thus, the correction for absorption is now by a factor of $\sim$3--4. Also note that there is a spread by a factor of $\sim$2 in the normalization of the spectra within each group. 

Interestingly, both Hj08a and SZ08 found relatively high fractions of nonthermal electrons in the hard state, which is in contrast to the hard state of black-hole binaries being usually dominated by thermal electrons, e.g., McConnell et al.\ (2002), ZG04. This may be an effect due to interactions of the very strong stellar wind with the inner part of the accretion flow in Cyg X-3. In our fits to the \xte\/ data, thermal Comptonization appears to dominate, and the limited high-energy sensitivity of HEXTE prevents us from obtaining precise constraints on nonthermal electrons. The intermediate and soft-state spectra are dominated by the disc component, but then followed by a significant hard tail clearly requiring the presence of both hot thermal electrons and nonthermal ones, similarly to the soft states of black-hole binaries, e.g., McConnell et al.\ (2002), ZG04.

\section{Radio emission and its correlation with X-rays}
\label{radio}

Here, we revisit the classification of the radio states of Waltman et al.\ (1994, 1995, 1996). We consider here the 8.3 GHz radio light curve together with those of 3--5 keV and 20--100 keV X-rays, see Fig.\ \ref{asmradio}. From joint analysis of the 8.3 GHz and 3--5 keV light curves, we have found the following six distinct variability patterns, with four of them corresponding to the previous classification. They are related to a transition level in both fluxes, shown by the horizontal dotted line in Fig.\ \ref{asmradio}. 

\begin{enumerate}

\item The quiescent state: both the radio and soft X-ray fluxes vary in a correlated manner below the transition level.

\item The minor-flaring state: the soft X-ray flux oscillates around its transition level while the radio flux varies up to its transition level. 

\item The suppressed state: the radio flux is below the transition level and the soft X-ray flux is above it, and the state is not directly followed by a radio flare.

\item The quenched state: the radio flux is much below the transition level and the soft X-ray flux is above it, and the state is followed by a major radio flare.

\item The major-flaring state: the soft X-ray flux is above the transition level, and the radio flux moves up and down by a large factor in a flare.

\item The post-flare state: the return after a major flare to either the minor-flaring state or the suppressed one. 

\end{enumerate}

\begin{table}
\caption{The numbered intervals of Fig.\ \ref{asmradio} assigned to the radio states. The colours given below are also used in Figs.\ \ref{asmradio2} and \ref{batse}.
}
\begin{center}
\begin{tabular}{ccc}
\hline
Radio state & Colour & Interval number\\
\hline
Quiescent    & red & 12, 14, 16, 18, 23\\
Minor flaring & blue & 19, 21, 24\\
Suppressed    & black & 6, 7, 8, 13, 15, 17, 20, 22\\
Quenched      & green & 3, 9, 25, 26, 28\\
Major flaring & magenta & 1, 4, 10, 27, 29\\
Post-flare    & cyan & 2, 5, 11 \\
\hline
\end{tabular}
\end{center}
\label{radio_states}
\end{table}

The transition level corresponds to the 3--5 keV count rate of $\simeq$3 s$^{-1}$, and the 8.3 GHz flux of $\simeq$0.3 Jy. It defines a boundary above which the positive correlation between the two fluxes breaks down. The radio transition level represents an approximate upper limit to the radio flux in most states, and it is crossed mostly during strong flares. The vertical solid lines in Fig.\ \ref{asmradio} divide intervals corresponding to different states. Most of the intervals are numbered, and Table \ref{radio_states} assigns them to the radio states. A few intervals that are not numbered correspond to states which we could not unambiguously classify.

\begin{figure}
\centerline{\includegraphics[width=\columnwidth]{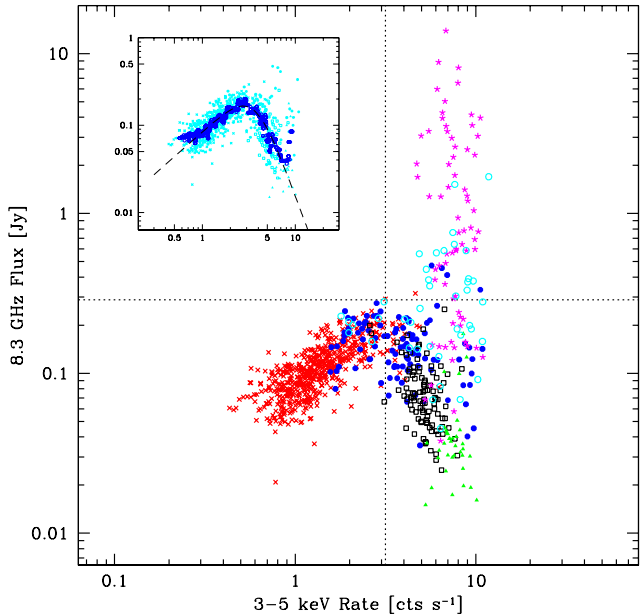}}
\caption{The 8.3 GHz radio flux as a function of the ASM 3--5 keV count rate. The red crosses, blue filled circles, black open squares, green filled triangles, magenta stars and cyan open circles correspond to the quiescent, minor-flaring, suppressed, quenched, major-flaring, and post-flare states, respectively, see Fig.\ \ref{asmradio} and Table \ref{radio_states}. The dotted lines represent the X-ray and radio transition levels shown in Fig.\ \ref{asmradio}. The inset shows the first four states only (cyan points), the 15-point running median of the points ordered according to the increasing ASM count rate (blue open circles) and a double power-law fit to the median points (dashed curve).
} 
\label{asmradio2}
\end{figure}

Fig.\ \ref{asmradio2} presents the correlation between the two fluxes for the intervals with the identified variability pattern, marked by different colours. The six different states form a clear sequence of correlated radio/X-ray behaviour, with the sequence of the quiescent, minor-flaring, suppressed, quenched, major-flaring, and post-flare states. 

\begin{figure}
\centerline{\includegraphics[width=\columnwidth]{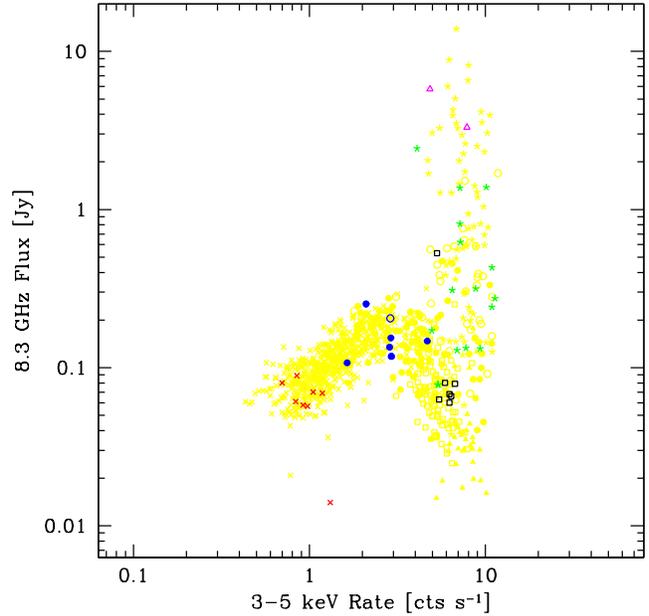}}
\caption{The GBI 8.3 GHz radio flux and the ASM 3--5 keV count rate measured on the same days as the pointed PCA/HEXTE observations shown in Fig.\ \ref{individual}. The red crosses, a blue open circle, black open squares, magenta open triangles, and green stars correspond to the X-ray spectral types 1, 2, 3, 4, 5, respectively, see Section \ref{xray}. The blue filled circles correspond to the observations in the X-ray type 2 for which GBI data were not available, and where we instead show the 15 GHz Ryle fluxes. The yellow background corresponds to all the data shown in Fig.\ \ref{asmradio2}. 
} 
\label{asmradio_pointed}
\end{figure}

Fig.\ \ref{asmradio_pointed} shows the location of the X-ray spectral states (see Fig.\ \ref{individual}) in the radio flux vs.\ the ASM X-ray flux diagram. Here, we mark the position of the ASM/radio points at the times closest in time to the pointed PCA/HEXTE observations. We see that the hardest X-ray state, type 1, corresponds to the quiescent radio state. Then, there were no GBI measurements corresponding to most of the X-ray spectral type 2. Thus, we have used instead the 15 GHz Ryle data (presented in Hj08a) in those cases. We clearly see that this X-ray state corresponds to the minor-flaring state. There were no pointed observations corresponding to the quenched state. Most of the pointed observations corresponding to the X-ray type 3 correspond to the suppressed state. Then the X-ray types 4 and 5 take place during the major-flaring state. The 3--5 keV fluxes of the pointed observations are in the $\sim (0.4$--$4)\times 10^{-9}$ erg cm$^{-2}$ s$^{-1}$ range. 

\begin{figure}
\centerline{\includegraphics[width=0.85\columnwidth]{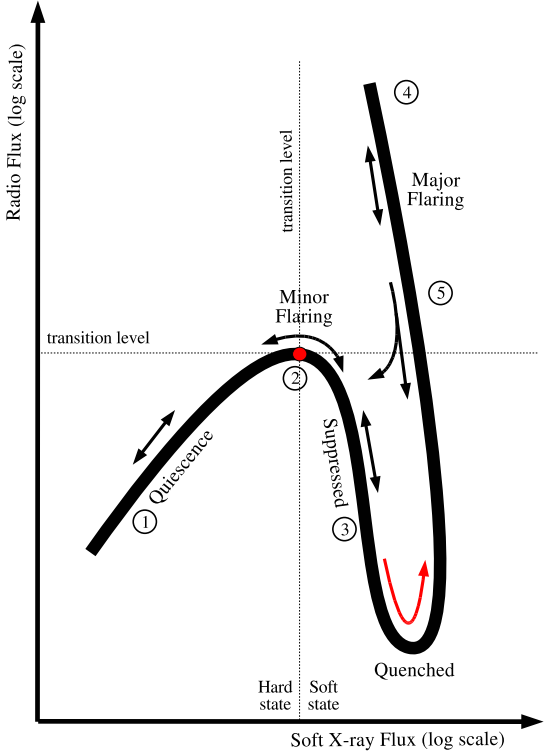}}
\caption{A schematic representation of the evolution of Cyg X-3 through its radio and X-ray states. The dotted lines represent the transition levels, with a dot at their intersection. The arrows show the possible directions of the source evolution. The numbers correspond (approximately) to the X-ray spectral states.}
\label{scheme}
\end{figure}

The sequence of radio and X-ray states we have found is shown schematically in Fig.\ \ref{scheme}. Its central point is shown by the dot, which corresponds to the crossing of the radio and X-ray transition levels of Fig.\ \ref{asmradio}. To the left of this point, Cyg X-3 is in the radio quiescent state and in the hard X-ray state (type 1). In the quiescent/hard state, the source moves up and down along this branch, and the radio emission is correlated with soft X-rays. 

The minor-flaring state (X-ray type 2) corresponds to a transition between the radio quiescent state and the suppressed state, as well as between the hard and soft X-ray states. These transitions may occur in both directions, without any apparent hysteresis effects (unlike black-hole transients, but similarly to Cyg X-1, ZG04). The radio/X-ray correlation changes its sign in this state from positive to negative. Consequently, a further increase of the X-ray flux corresponds to a decrease of the radio flux, first to the suppressed state and then to the quenched one. As mentioned above, the suppressed state (which we introduce in this work) is not followed by a radio flare. From this state, the source has still been observed to return to the quiescent/hard state through the minor-flaring one, see Fig.\ \ref{asmradio}. 

If, however, the X-ray and radio fluxes keep increasing and decreasing, respectively, Cyg X-3 enters the quenched state. At this point, the 3--5 keV and 8.3 GHz fluxes achieve their overall maximum and minimum, respectively. In both the suppressed and quenched states, the radio emission is anticorrelated with the 3--5 keV X-rays. 

Interestingly, all major radio flares have been observed only after a radio quenched period, and in all cases except one (the interval 25 in Fig.\ \ref{asmradio}, in which case after some initial increase of the radio flux the source returned to the quenched state, interval 26), the quenched state is followed by a major flare. While on the major-flaring branch, the radio flux changes by more than three orders of magnitude while at the same time the 3--5 keV flux remains almost constant, and the X-ray spectra belong to the types 4, 5, see Fig.\ \ref{asmradio_pointed}.  

The decline of the radio flux after a flare is through the minor-flaring and suppressed states, as shown in Fig.\ \ref{asmradio2}, but {\it not\/} through the quenched state. Thus, we see a hysteresis in this type of transitions, with the source following a loop in the flux-flux diagram, see Fig.\ \ref{scheme}. The post-flare stage is not strictly a new state but it is defined by its temporal appearance after a major flare.

We now quantify the radio/soft X-ray correlation in the first four radio states, shown by the cyan points in the inset in Fig.\ \ref{asmradio2}. The correlation changes its sign from positive to negative, which can be fitted. However, in order to model the dominant long-term trend and reduce the scatter, we first generate the 15-point running median from those points ordered according to the increasing ASM flux. As shown by the blue points in the inset in Fig.\ \ref{asmradio2}, this indeed strongly reduces the scatter and yields a narrow path on the diagram. We then fit a a double power law function,
\begin{equation}
F_{\rm R}={1\over A_1^{-1} F_{\rm X}^{-p_1}+A_2^{-1} F_{\rm X}^{-p_2}},
\label{fit1}
\end{equation}
which has an asymptotic power-law form, $A_i F_{\rm X}^{p_i}$, on either side of the break, and where $F_{\rm R}$ and $F_{\rm X}$ is the radio flux and the X-ray count rate, respectively. We find the best fit of $A_1=0.085\pm 0.001$ Jy, $p_1=0.95\pm 0.02$, $A_2=12.5\pm 2.0$ Jy, $p_2=-2.91\pm 0.09$. We have also calculated the Spearman rank correlation coefficients for the states (i) and (iii)--(iv), which are $+0.75$, $-0.51$, with the null hypothesis (of no correlation) probability of $\ll 10^{-10}$, $\sim 10^{-9}$, respectively.

\begin{figure}
\centerline{\includegraphics[width=8cm]{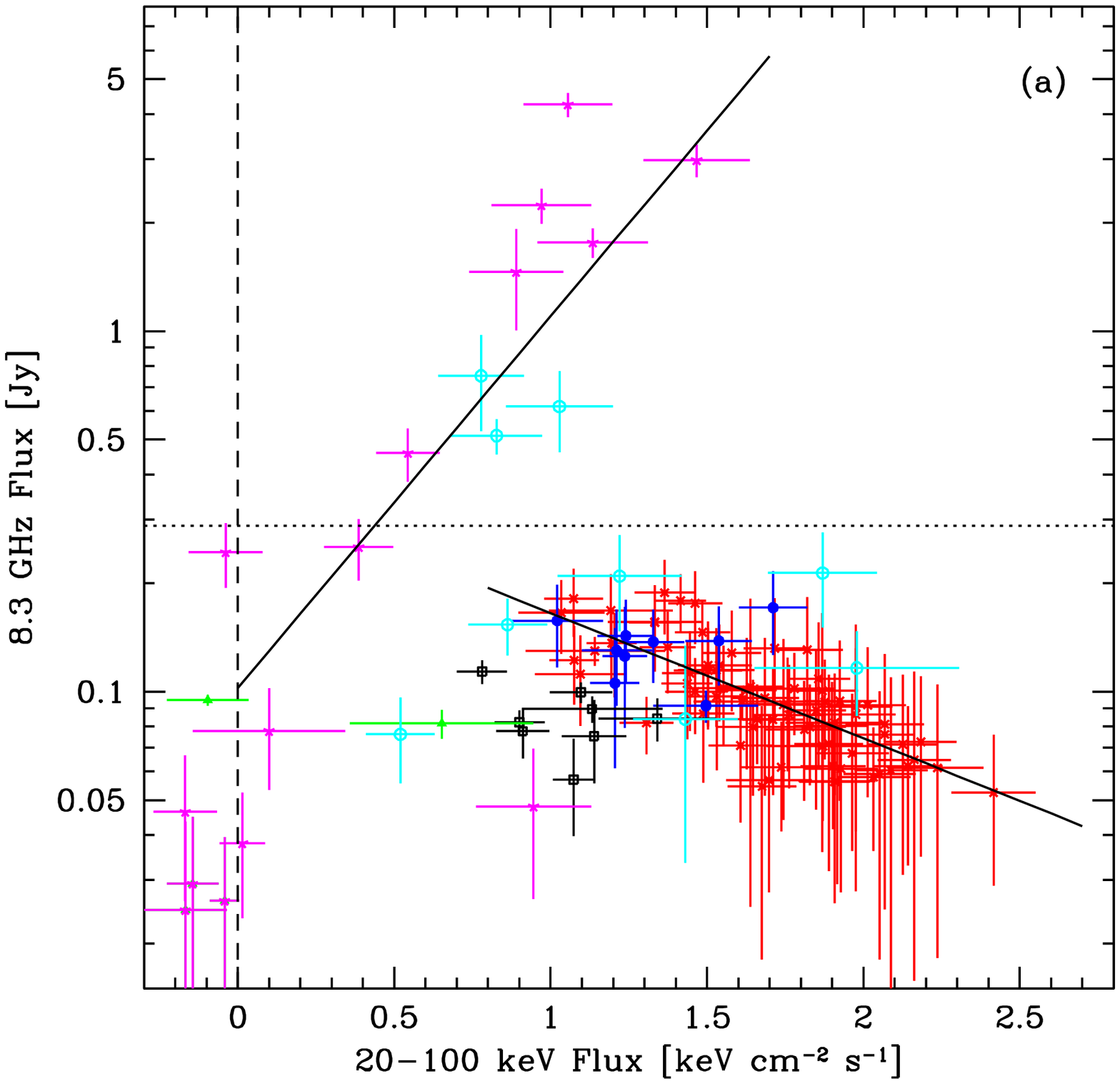}}
\centerline{\includegraphics[width=8cm]{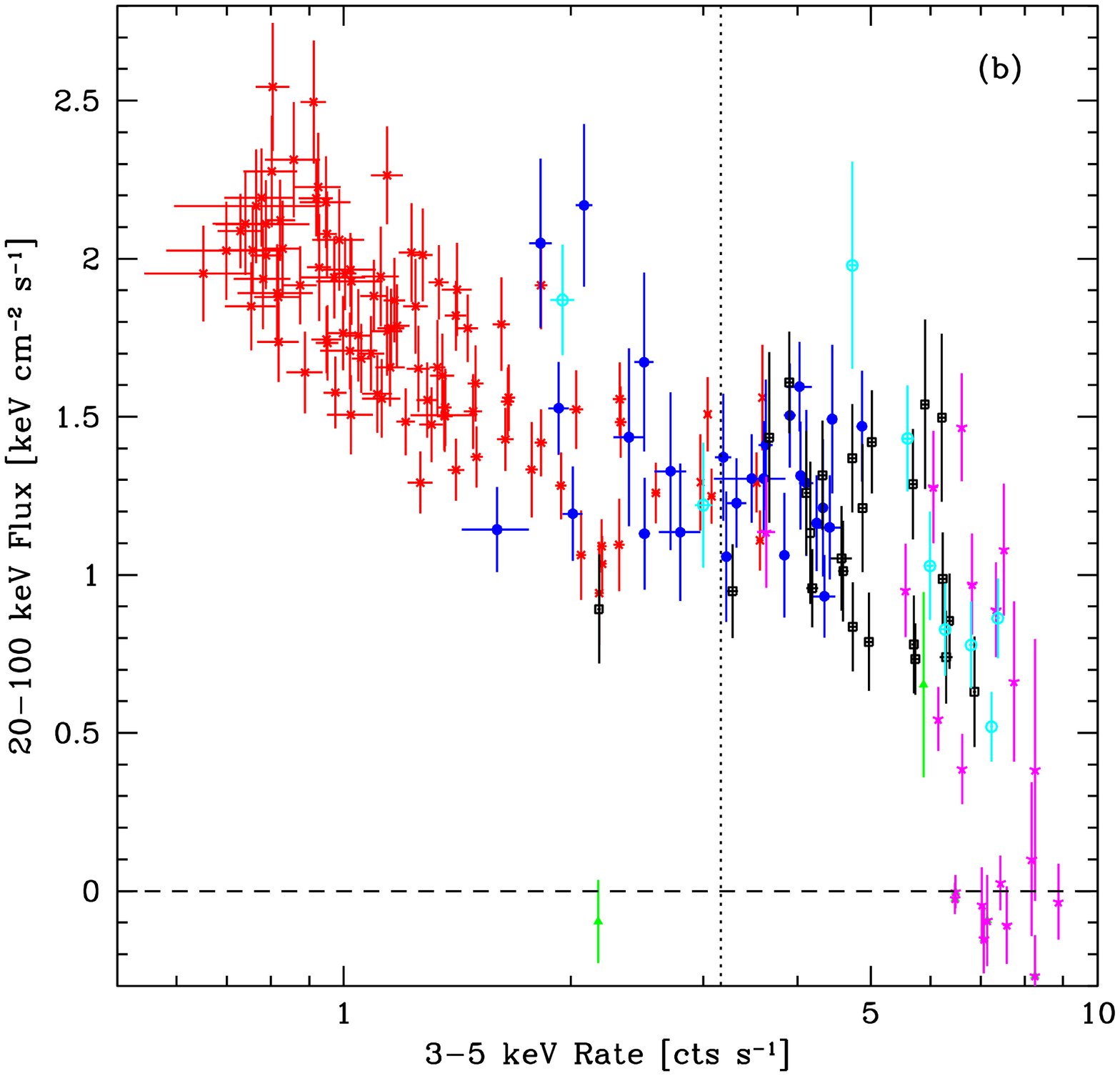}}
\caption{(a) The  GBI flux as a function of the BATSE flux. On both panels, the colours correspond to the radio states, see Fig.\ \ref{asmradio2} and Table \ref{radio_states}, the dashed line gives the zero BATSE flux and the dotted line corresponds to the transition levels defined in Fig.\ \ref{asmradio}. The solid lines give fits to the anticorrelation in the quiescent and minor flaring states, and to the positive correlation in the major-flaring state. (b) The BATSE flux as a function of the ASM flux.
}
\label{batse} 
\end{figure}

We now compare the radio/soft X-ray correlation with that for hard X-rays. Fig.\ \ref{asmradio} shows the 20--100 keV BATSE light curve. Fig.\ \ref{batse}(a) shows the 8.3 GHz flux as a function of the 20--100 keV flux. We see an anticorrelation in the quiescent/hard state, which changes into a positive correlation on the major-flaring branch (including a few points from the post-flare state). This behaviour is the same as that found by McCollough et al.\ (1999) for an earlier data set; however, we now show the position of the radio states on the flux-flux diagram. The two correlations form two clear branches. We have fitted them by a straight line on the linear-logarithmic plot, namely,
\begin{equation}
F_{\rm R}= F_{\rm R,i}\exp\left(F_{\rm HX}/ F_{\rm HX,i}\right),
\label{fit2}
\end{equation}
where $F_{\rm R,i}$ is the radio flux corresponding to the null hard X-ray flux, $F_{\rm HX}$, and $F_{\rm HX,i}$ is the e-folding flux. We obtain $F_{\rm R,1}=0.37\pm 0.03$ Jy, $F_{\rm HX,1}= -1.25\pm 0.09$ keV cm$^{-2}$ s$^{-1}$ for the quiescent branch, and $F_{\rm R,2}=0.10\pm 0.08$ Jy, $F_{\rm HX,2}= 0.42\pm 0.13$ keV cm$^{-2}$ s$^{-1}$ for the major-flaring branch (but including the three post-flare points with the fluxes above the transition level, see Fig.\ \ref{batse}a). Thus, on the major-flaring branch, the radio increases exponentially with an increase of the hard X-ray flux. The overall evolutionary track is from the quiescent state to the minor flaring, where the negative correlation already changes sign, and both the radio and the hard X-ray fluxes drop to the quenched state. Then, they increase again along the positive correlation, see Fig.\ \ref{batse}(a). The return from a major flare avoids the quenched state and it goes directly to the minor flaring (or, sometimes, suppressed) state, as shown by the cyan points. 

Fig.\ \ref{batse}(b) shows the corresponding ASM/BATSE diagram. There is a strong anticorrelation in the hard/quiescent state, after which the 20--100 keV flux becomes independent of the 3--5 keV one until the major-flaring state, in which the correlation becomes again negative and the 20--100 keV flux drops to zero. 

\begin{figure}
\centerline{\includegraphics[width=8cm]{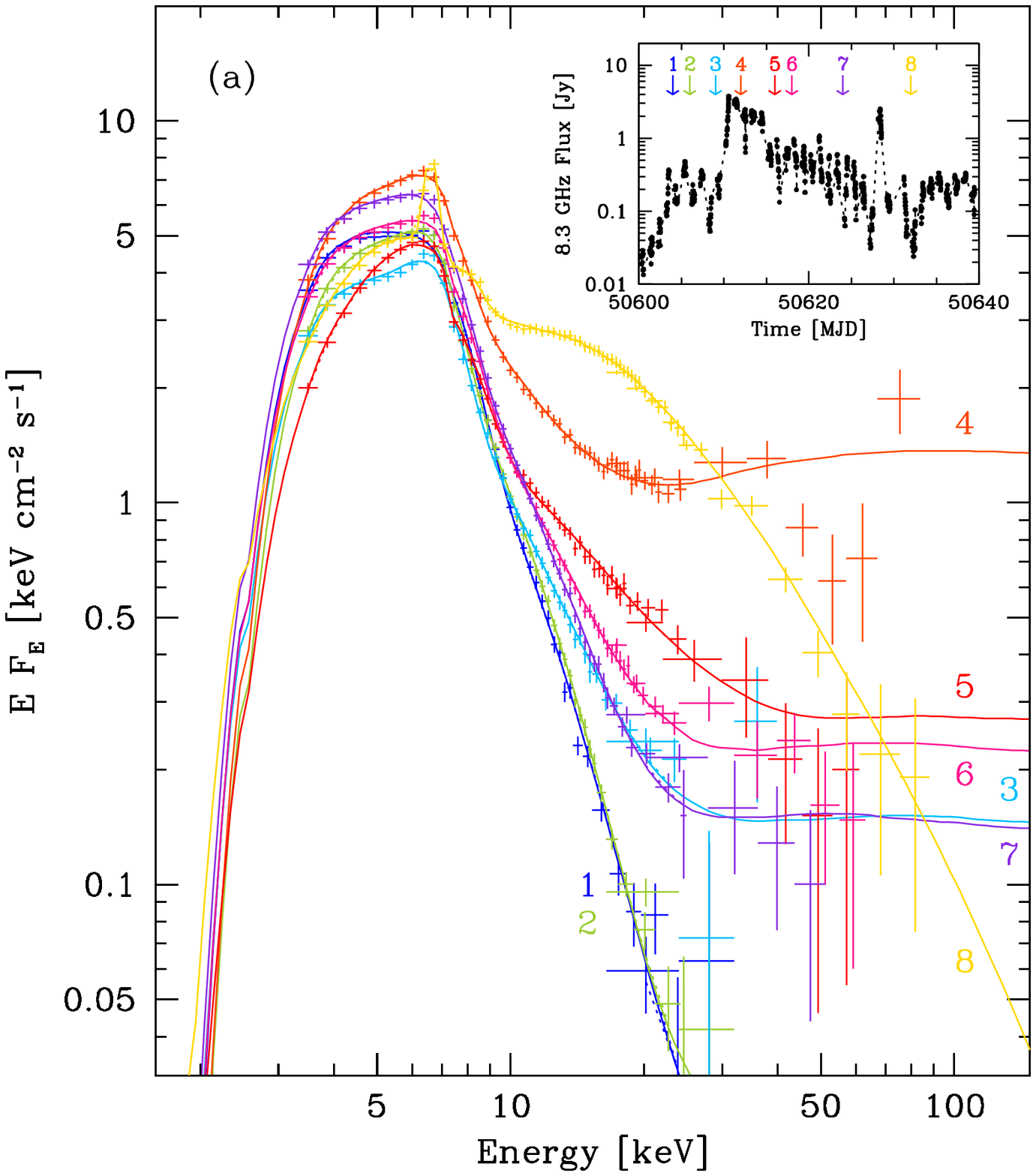}}
\centerline{\includegraphics[width=8cm]{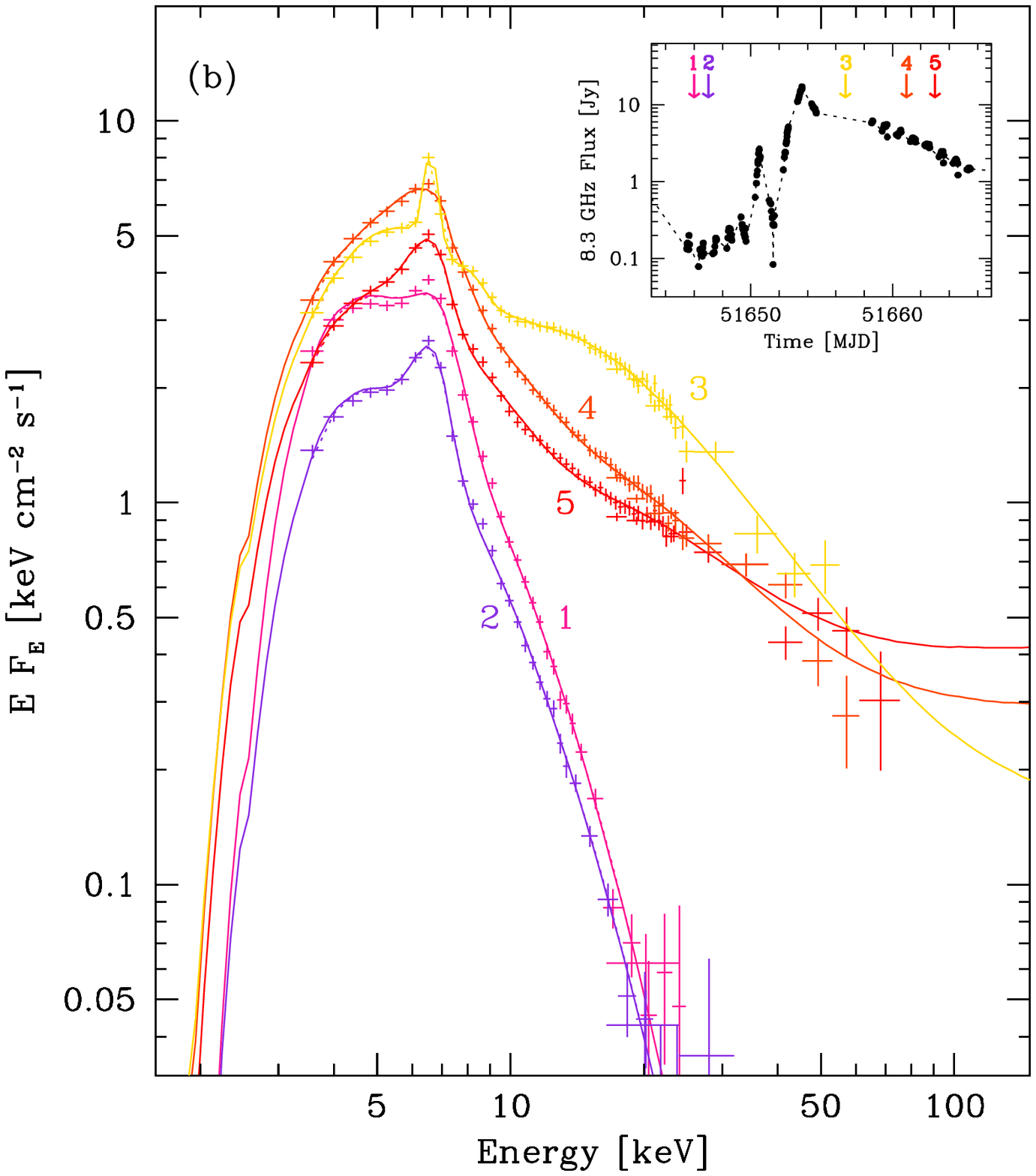}}
\caption{The PCA/HEXTE spectra taken during two major radio flares. The numbers correspond to the increasing time, and are identified in the insets, giving the radio light curves. }
\label{majorflare} 
\end{figure}

Fig.\ \ref{majorflare} presents two series of the PCA/HEXTE spectra taken during the radio flares that started at MJD $\sim$50610 and $\sim$51650. The spectra were fitted with the same model as in Section \ref{xray}. We see that the hard tail flux varies much more than that of the soft X-rays. The flares start with ultrasoft spectra with no high-energy tail and then the tail strength increases, going through a sequence of the X-ray spectral types 5, 4, 3. Thus, the strength of the high-energy tail is positively correlated with the radio flux, as we have already seen in Fig.\ \ref{batse}(a).

\section{Discussion and comparison to other X-ray binaries}
\label{discussion}

In Section \ref{xray}, we have presented a detailed classification of the X-ray spectra of Cyg X-3, forming 5 types defined by the spectral hardness and the strength of the soft, disc-like component. The average spectrum of each type was well fitted by a physical model including Comptonization of disc blackbody photons by both thermal and nonthermal electrons, Compton reflection and an Fe K line. 

The correspondence of these spectral types to the canonical X-ray states of black-hole binaries was illustrated in fig.\ 9 of ZG04. There, the types 1--2 were classified as low/hard, the type 3 as intermediate/very high, the type 4 as high, and the type 5 as ultrasoft, and found to closely correspond to the spectral states of the black-hole transient XTE J1550--564. Also, the intermediate and soft spectra of Cyg X-3 are rather similar to those of GRS 1915+105 (Zdziarski et al.\ 2001, 2005). In the softest state, we have found a distinct high-energy tail with the photon index of $\Gamma\simeq 2$ extending above 100 keV. Such a spectral feature also appears in the ultrasoft states of some other black-hole binaries, in particular GRS 1915+105 (Zdziarski et al.\ 2001) and XTE J1550--564 (ZG04). The origin of this tail is probably due to Comptonization by nonthermal electrons. The main difference in the spectra of Cyg X-3 and those of black-hole binaries is the high-energy break in the hard state, which occurs in Cyg X-3 at $\sim$20 keV whereas it is at $\ga$100 keV in other objects, see, e.g., ZG04. As discussed in Hj08a and SZ08, a possible origin of this difference is the interaction of the very strong stellar wind in Cyg X-3 with a hot inner flow, but there is no quantitative model as yet.

On the other hand, there is also a certain similarity between the X-ray spectra of Cyg X-3 and atoll sources (which are weakly-magnetized neutron-star low-mass X-ray binaries), see, e.g., the spectra of 4U 1608--52 in Gierli\'nski \& Done (2002). However, the spectra of Cyg X-3 correspond to significantly higher luminosities than those of atoll sources (Hj08a, SZ08). Also, as noticed in SZ08, all neutron stars in confirmed high-mass X-ray binaries are highly magnetized, and the spectra of those bear no resemblance to Cyg X-3. Thus, the form of X-ray spectra of Cyg X-3 points out to (though does not prove) the presence of a black hole in that system.

In Section \ref{radio}, we have studied in detail the correlated behaviour of the radio emission and soft and hard X-rays in all states of Cyg X-3. Based on the correlated behaviour of radio and soft X-rays, we have identified six states, see Table \ref{radio_states}. Five of them correspond to adjacent regions along a path on its flux-flux diagram. The evolution can go in either direction on that path except for the weakest radio state, the quenched one, which almost always leads to a major radio flare. The sixth radio state corresponds to the decline of major radio flares, which follows along paths avoiding the quenched state, thus forming a hysteresis loop. 

The radio/soft X-ray correlation is positive in the quiescent radio state, which corresponds to the hard spectral state (the type 1), and changes its sign to negative within the minor-flaring radio state, which corresponds to a softening of the X-ray spectra. The radio flux reaches its minimum when the soft X-ray flux is at its maximum. This point is followed by a major radio flare, when the radio flux can increase by $\sim\! 10^3$ at an approximately constant soft X-ray (3--5 keV) flux.

We find that overall the radio/soft X-ray correlation in Cyg X-3 is rather similar to those of black-hole binaries. In the hard state, the positive correlation in Cyg X-3 is very similar to that of black-hole binaries in the hard state (Corbel et al.\ 2000, 2004; GFP03; Fender, Belloni \& Gallo 2004). When Cyg X-3 enters softer states from the hard state, the radio flux starts to decline and the correlation changes its sign. This is again similar to other black hole binaries. In particular, the radio flux strongly declines at the transition from the hard state to a soft one in Cyg X-1 (e.g., GFP03), GX 339--4 (Corbel et al.\ 2000), XTE J1650--500 (Corbel et al.\ 2004), and XTE J1859+226 (Fender et al.\ 2004). After the decline, the radio flux from black-hole transients strongly increases, sometimes up to the overall maximum for a source, see Fender et al.\ (2004). This behaviour is also seen in the black-hole sample shown in fig.\ 7 of Merloni, Heinz \& Di Matteo (2003). A very bright radio flare of GX 339--4 took place during a transition from the hard state to a soft (very-high) one (Gallo et al.\ 2004), which typifies the above pattern. Although Cyg X-1, a persistent source, does not generally follow the whole of the above pattern (which appears to be due to the limited range of luminosities it covers), it also shows episodes of a radio flux in its soft state as strong as in the hard state, see, e.g., its radio light curve in Lachowicz et al.\ (2006). The two brightest observed radio flares of Cyg X-1 happened in intermediate/soft spectral states (Fender et al.\ 2006; Wilms et al.\ 2007). As mentioned above, GRS 1915+105 is never in the hard state, and its radio/X-ray flux evolution resembles the quenched/major-flare branch of Cyg X-3, see Klein-Wolt et al.\ (2002), GFP03. All of those similarities provide a very strong argument for the black-hole nature of the compact object in Cyg X-3.

We note that the normalization of the radio flux in the hard state for a given soft X-ray flux is higher by an order of magnitude in Cyg X-3 than in other black-hole binaries (GFP03). A part of this effect is clearly due to the strong X-ray absorption in Cyg X-3, e.g., compare the absorbed and unabsorbed hard state X-ray spectra in SZ08. Still, this may account only for about a factor of two in the \xte/ASM range of 1.5--12 keV. To avoid the effect of X-ray absorption, we may consider a correlation of the radio flux with the unabsorbed bolometric X-ray fluxes. Such correlation was calculated for GX 339--4 by Zdziarski et al.\ (2004), who obtained 
\begin{equation}
F(8.6\,{\rm GHz})/1\,{\rm mJy}\simeq 2.8\pm 0.3 (F_{\rm bol}/1\,{\rm keV\, cm}^{-2}\, {\rm s}^{-1})^{0.79\pm 0.07}.
\label{fit3}
\end{equation}
For the our hard-state sample, the average radio flux is about 60 mJy, whereas the bolometric X-ray flux corrected for absorption is 10 keV cm$^{-2}$ s$^{-1}$ ($1.6\times 10^{-8}$ erg cm$^{-2}$ s$^{-1}$), see Hjalmarsdotter et al.\ (2008b). The above relation implies then the radio flux of only 17 mJy, which means that Cyg X-3 in the hard state for a given bolometric X-ray flux is $\sim$3--4 times overluminous in radio compared to GX 339--4. Some reduction of this factor may be due to Cyg X-3 being possibly seen close to edge-on, which is likely to result in the X-ray emission being partly obscured in our direction (see Hj08a) as well as moderately beamed toward the polar direction, but this is unlikely to account for this factor of three. The net overluminous radio emission may be due to a turbulent interaction of the very strong stellar wind with the hot inner accretion flow, resulting in copious acceleration of electrons to nonthermal energies. (As discussed above, the same effect may be responsible for the peculiar spectral shape of the X-ray spectrum in the hard state.) 

On the other hand, weakly-magnetized neutron-star X-ray binaries produce much less, by an order of magnitude, radio power at a given soft X-ray flux than the black-hole binaries (Migliari \& Fender 2006). Also, the slope of hard-state correlation is similar in Cyg X-3 and in other black hole binaries (GFP03) whereas it is much steeper, with the power-law index $\ga$1.4, in neutron-star binaries (Migliari \& Fender 2006). Thus, the presence of a neutron star in Cyg X-3 appears highly unlikely.

We have also studied correlations involving hard X-rays. Overall, the hard X-rays, 20--100 keV in our study, are anticorrelated with the 3--5 keV X-rays, see Fig.\ \ref{batse}(b), which corresponds to spectral pivoting with a pivot energy between the two bands (see Figs.\ \ref{individual}--\ref{groups}, also Choudhury et al.\ 2002). This behaviour is again similar to that of black-hole binaries, in particular in Cyg X-1, where a strong correlation between the 1.5--3 keV ASM and 100--300 keV BATSE fluxes is seen, see fig.\ 9b of Zdziarski et al.\ (2002). Note that the blackbody temperature in Cyg X-3 is higher (Hj08a; SZ08) than that in Cyg X-1, so the 3--5 keV band is a better indicator of the disc flux in the former. The hard-state spectrum of Cyg X-3 has a significantly lower cutoff energy than that typical for black-hole binaries, so the 20--100 keV band samples the spectral region of the cutoff in Cyg X-3, whereas the 100--300 keV band does it in Cyg X-1. After those adjustments, the hard/soft X-ray anticorrelations in the hard state in Cyg X-1 and Cyg X-3 appear very similar. The likely physical explanation of it in Cyg X-1 is a variable flux in seed photons irradiating the hot Comptonizing plasma supplied with an approximately constant power, which results in the plasma temperature decreasing with the increasing seed photon flux (see fig.\ 14 in Zdziarski et al.\ 2002). A similar explanation is likely to hold in Cyg X-3. Also, in both objects, the BATSE flux drops to very low values in the soft states with the highest ASM flux. Both objects also show a flat dependence in intermediate soft states. 

Given the hard-state anticorrelation between the soft and hard X-rays, the positive correlation of the radio flux with the soft X-rays becomes a negative one with the hard X-rays, see Fig.\ \ref{batse}(a). The fact that the correlation of the radio emission is positive with the soft X-rays points out to the importance of the blackbody-emitting disc (or its inner edge) to the jet formation in the hard state.

On the other hand, while the soft X-ray flux is almost constant during major flares, the hard X-ray flux is strongly (and positively) correlated with the radio flux during major flares. The X-ray spectra during the major flares are soft/high, very high, or ultrasoft (Fig.\ \ref{majorflare}). The dominant spectral components during major flares in the 3--5 keV and 20--100 keV bands are the weakly varying (disc) blackbody and a strongly varying high-energy tail, respectively, see Fig.\ \ref{majorflare}. This behaviour is likely explained by the presence of a stable accretion disc and variable active regions above the disc, and it is again typical to black-hole binaries, see the analogous result for the soft state of Cyg X-1 in Zdziarski et al.\ (2002). In the case of Cyg X-3, we have discovered that the radio flux during major flares responds exponentially to the amplitude of the high-energy tail, see Fig.\ \ref{batse}(a) and equation (\ref{fit2}). This shows that the radio-emitting region communicates, on time scales of the order of a day, with the X-ray source, and it is not detached from it. 

An important issue is the origin of the high-energy, hard X-ray, tail in the soft states. Gierli\'nski et al.\ (1999), Zdziarski et al.\ (2001, 2002, 2005) and Gierli\'nski \& Done (2003) modelled this tail in the black-hole binaries Cyg X-1, GRS 1915+105, and XTE J1550--564 by Comptonization in a hybrid plasma, in which the electron distribution contains a significant non-thermal high-energy tail. This Comptonization presumably takes place in some coronal regions above the accretion disc. The correlation of the tail amplitude with the radio flux may be then due to the formation of radio-flare emitting blobs being related to the non-thermal coronal regions. Still, the observed radio flux has to be due to synchrotron emission in jet regions relatively far away from the compact object, which is required by the lack of any orbital modulation in Cyg X-3 by free-free absorption in the stellar wind (Szostek \& Zdziarski 2007). 

On the other hand, a rather common model in the astrophysical literature attributes hard X-ray tails of black-hole binaries in soft states directly to emission of non-thermal electrons in the jet, usually via the synchrotron process. There are two major problems with this model (see Zdziarski et al.\ 2003). One is that the spectra of the tails are generally soft, with the photon indices of $\Gamma\ga 2$, typically $\sim$3, which is also the case in most of the soft states of Cyg X-3 (Fig.\ \ref{groups}). Such a soft spectrum extrapolated to the turnover frequency, typically in the infrared, would yield a huge luminosity, orders of magnitude higher than those observed. Second, the high-energy tails are always smoothly connected to the disc component, which is naturally explained by the tails being due to Comptonization of the disc photons. If they were due to a process unrelated to the disc, their amplitude in the hard X-ray region could be any, without any reason for this smooth connection. The first problem can be circumvented by requiring that the synchrotron spectrum is hard at low energies and it breaks to a soft one just below the region where the tails are observed. This, however, requires yet another fine-tuning, and given the large observed sample of high-energy tails, can be ruled out on statistical grounds.

It is of great interest whether the non-thermal electrons responsible for either the hard X-ray tails or the radio emission during major flares may be accelerated to high enough energies to result in detectable emission in the $\gamma$-ray range, e.g., the GeV or TeV band. Up to now, no TeV emission from Cyg X-3 has been confirmed (compare, e.g., Lamb et al.\ 1982 with Weekes 1983). However, given our result on major radio flares being indicators of the presence of hard X-ray tails, GeV and TeV emission should be searched for during those radio flares.

Finally, we point out that the radio source in Cyg X-3 consists of the core and the jet components, which can be resolved (e.g., Mart{\'{\i}}, Paredes \& Peracaula 2000; Mioduszewski et al.\ 2001). In fact, the e-VLBI results of Tudose et al.\ (2007) show that the core may become undetectable during a radio flare. On the other hand, our analysis in the radio band has been done using data from the GBI, which beam includes both the core and jet components, which did not allow us to include the spatial dimension in the correlation analysis. Also, our analysis did not take into account time lags of the radio emission with respect to the X-rays. These issues, as well as development of theoretical models, should be dealt with in future studies of radio/X-ray correlations. 

\section*{Acknowledgments}
We thank L. Hjalmarsdotter for valuable discussions. This research has been supported in part by the Polish MNiSW grant NN203065933 (2007--2010) and the Polish Astroparticle Network 621/E-78/SN-0068/2007. The GBI was operated by the National Radio Astronomy Observatory for the U.S.\ Naval Observatory and the Naval Research Laboratory during the the observations used here. We acknowledge the use of data obtained through the HEASARC online service provided by NASA/GSFC.

\label{lastpage}

\end{document}